\documentclass[prc,twocolumn,floatfix,superscriptaddress,showpacs]{revtex4}
\usepackage{amssymb,graphicx,bm}
\usepackage{dcolumn}

\newcommand{\mev}{\, \text{MeV}}
\newcommand{\gcq}{\, \text{g}/\text{cm}^{3}}

\begin{document}

\title{Influence of light nuclei on neutrino-driven supernova outflows}

\author{A.~Arcones}
\affiliation{Institut f\"ur Kernphysik, TU~Darmstadt,
Schlossgartenstr.~9, D-64289 Darmstadt, Germany}
\affiliation{Gesellschaft f\"ur Schwerionenforschung,
Planckstr.~1, D-64291 Darmstadt, Germany}
\author{G.~Mart\'inez-Pinedo}
\affiliation{Gesellschaft f\"ur Schwerionenforschung,
Planckstr.~1, D-64291 Darmstadt, Germany}
\author{E.~O'Connor}
\affiliation{TRIUMF, 4004 Wesbrook Mall, Vancouver, BC,
V6T 2A3, Canada}
\affiliation{Department of Physics, California Institute of
  Technology, Pasadena, CA 91125}    
\author{A.~Schwenk}
\affiliation{TRIUMF, 4004 Wesbrook Mall, Vancouver, BC,
V6T 2A3, Canada}
\author{H.-Th.~Janka}
\affiliation{Max-Planck-Institut f\"ur Astrophysik,
Karl-Schwarzschild-Str.~1, D-85741 Garching, Germany}
\author{C.J.~Horowitz}
\affiliation{Nuclear Theory Center and Department of Physics,
Indiana University, Bloomington, IN 47408}
\author{K.~Langanke}
\affiliation{Gesellschaft f\"ur Schwerionenforschung,
Planckstr.~1, D-64291 Darmstadt, Germany}
\affiliation{Institut f\"ur Kernphysik, TU~Darmstadt,
Schlossgartenstr.~9, D-64289 Darmstadt, Germany}


\begin{abstract}
  We study the composition of the outer layers of a protoneutron star
  and show that light nuclei are present in substantial amounts. The
  composition is dominated by nucleons, deuterons, tritons and alpha
  particles; $^3$He is present in smaller amounts. This composition
  can be studied in laboratory experiments with new neutron-rich
  radioactive beams that can reproduce similar densities and
  temperatures. After including the corresponding neutrino
  interactions, we demonstrate that light nuclei have a small impact
  on the average energy of the emitted electron neutrinos, but are
  significant for the average energy of antineutrinos. During the
  early post-explosion phase, the average energy of electron
  antineutrinos is slightly increased, while at later times during the
  protoneutron star cooling it is reduced by about $1 \mev$. The
  consequences of these changes for nucleosynthesis in neutrino-driven
  supernova outflows are discussed.
\end{abstract}

\pacs{26.50.+x, 26.30.-k, 25.30.Pt, 97.60.Bw}

\maketitle

\section{Introduction}
\label{sec:introduction}

Core-collapse supernovae are the birth places of neutron
stars. During the explosion, around $10^{53} \, \text{ergs}$,
corresponding to the binding energy of the neutron star, are emitted
in all neutrino species from a thermal surface denoted as
neutrinosphere. The spectrum and luminosities of the neutrinos
radiated from the newly formed neutron star, the so-called
protoneutron star, are of great importance for the interpretation of
supernova neutrino detections, for studies of flavor conversion
in the star
mantle~\cite{Lunardini.Mueller.Janka:2007,Duan.Fuller.ea:2008},
and for nucleosynthesis occurring in the so-called neutrino-driven
wind~\cite{duncan.shapiro.wasserman:1986,Qian.Woosley:1996}.

A crucial input for protoneutron star evolution is the equation of
state (EOS). Currently available simulations use either the
Lattimer and Swesty~\cite{Lattimer.Swesty:1991} or Shen \emph{et
al.}~\cite{Shen.Toki.ea:1998a} EOS, or an EOS based on 
nuclear statistical equilibrium 
(NSE)~\cite{Arcones.Janka.Scheck:2007}. All these EOS 
describe the nucleonic composition by a mixture of
neutrons, protons, alpha particles and a representative heavy nucleus.
While these EOS include the essential composition for the relatively
low densities and high temperatures present in the protoneutron star
atmosphere, where matter is fully dissociated,
they cannot account for the composition
in the region of the crust where the larger densities ($\rho \sim
10^{12} \gcq$) allow for the formation of light nuclei with
$A=2$ and $3$ in addition to alpha
particles~\cite{oconnor.gazit.ea:2007,sumiyoshi.roepke:2008}. In
contrast to the conditions in cold neutron stars, the high
temperatures ($T \gtrsim 4 \mev$) of the protoneutron star crust
suppress the formation of heavier nuclei. Deeper in the
protoneutron star the densities become so large that light nuclei
melt and a transition from inhomogeneous phases to 
homogeneous nuclear matter takes
place~\cite{Baym.Bethe.Pethick:1971,Schmidt.Roepke.Schulz:1990}.
As we will discuss below,
the change in composition caused by the presence of light nuclei in
the outer crust affects the neutrino opacities and
consequently changes their spectra and luminosities. In addition,
light nuclei are present in the region behind the shock, where 
the emitted neutrinos are expected to deposit 
their energy in the delayed supernova explosion
mechanism~\cite{Bethe.Wilson:1985,Janka.Langanke.ea:2007}, and
therefore interactions with neutrinos have to be included as
well~\cite{oconnor.gazit.ea:2007,sumiyoshi.roepke:2008,%
Ohnishi.Kotake.Yamada:2007}.

The neutrinos emitted by the young protoneutron star produce an
outflow of baryonic matter known as the neutrino-driven wind 
that has been the subject of many studies including full hydrodynamical
simulations~\cite{Arcones.Janka.Scheck:2007},
analytical~\cite{Qian.Woosley:1996} and 
steady-state approaches (see Ref.~\cite{Thompson.Burrows.Meyer:2001}
and references therein). This outflow is
initially very hot and essentially consists of free neutrons and protons in a
ratio that is determined by the competition of neutrino and
antineutrino absorptions on nucleons and their inverse reactions.
But as the matter expands and cools, nucleons can be
assembled into nuclei, and elements even heavier than iron can be
formed. If this occurs with a large abundance of free neutrons
present, these can be captured on heavy nuclei (the ``seed'')
producing an
r-process~\cite{Meyer.Mathews.ea:1992,Woosley.Hoffman:1992}. For a
successful r-process a large neutron-to-seed ratio is necessary,
requiring outflows with short dynamical time scales (a few
milliseconds), high entropies (above $150 \, k_{\text{B}}$) and low
electron fractions ($Y_e < 0.5$)~\cite{hoffman.woosley.qian:1997}.

In the present paper, we explore the influence of light
nuclei on the spectra and luminosities of electron neutrinos and
antineutrinos emitted during the cooling phase of the protoneutron
star and consequently on the electron-to-baryon ratio of the ejected
matter. The existence of light nuclei potentially also affects
$\mu$ and $\tau$ neutrinos. We plan to investigate this in future
work. The early post-bounce evolution and pre-explosion phase might also be
affected by the presence of light nuclei in the matter composition. However,
this issue cannot be explored here in detail, because 
the neutrino transport conditions in those phases and the
development of hydrodynamical instabilities in the forming neutron
star and in the neutrino heating region (for reviews, see 
Refs.~\cite{Burrows.Dessart.ea:2007,Janka.Langanke.ea:2007}) require
full radiation hydrodynamics simulations, which are beyond our study here.

If light nuclei are present in the region near the neutrinospheres
they will influence neutrinos and antineutrinos differently.  Deep in
the interior of the protoneutron star neutrinos, are in chemical
equilibrium with matter. At larger radii the decrease of temperature
and density allows for neutrinos to decouple from matter near the
protoneutron star surface and near the region where nuclei form.  For
the very neutron-rich conditions present in this environment, the
formation of light nuclei occurs mostly at the expense of free
protons. As protons represent the major source of opacity for electron
antineutrinos, and because the antineutrino cross sections on protons
and light nuclei are different for the relevant energies, light nuclei
can potentially affect the spectra and luminosities of the emitted
antineutrinos. The situation is not the same for electron neutrinos.
The abundance of neutrons is so large that it is insignificantly
changed by the appearance of light nuclei.  Consequently, the electron
neutrino opacity, which is dominated by interactions with neutrons,
remains practically unchanged. This asymmetrical effect on the
radiated neutrinos and antineutrinos can potentially change the
proton-to-neutron ratio of the ejecta and consequently the
nucleosynthesis. A study of these effects requires an EOS that
includes light nuclei in the composition and their corresponding
neutrino cross sections.

This warm nuclear system near the neutrinosphere, including the abundances
of light nuclei, can be studied with heavy-ion collisions. Fragments
emitted from a system at intermediate velocities may come from a 
low-density region between the colliding nuclei. The density and 
temperature of this region can be similar to conditions near the
neutrinosphere. Kowalski {\it et al.}~\cite{Kowalski:2006ju}
have measured the abundances of light nuclei (deuterons, tritons, 
$^3$He, and $^4$He) in near Fermi energy heavy-ion collisions of
$^{64}$Zn on $^{92}$Mo and on $^{197}$Au. They found $^4$He 
abundances and symmetry energies similar to those predicted by the virial 
EOS~\cite{horowitz.schwenk:2006}. In the future, experiments with
radioactive beams will enable studies of the more neutron-rich
neutrinosphere conditions.

This paper is organized as follows.
Section~\ref{sec:equat-state-opac} presents the EOS
(which we use for the determination of light-element abundances),
the neutron star atmosphere model, and the neutrino cross sections with light
nuclei. Section~\ref{sec:neutr-lumin-spectra} discusses the impact of
light nuclei on the spectra and luminosities of electron neutrinos and
antineutrinos and their influence on the electron fraction of the ejecta.
Finally, we conclude in Sect.~\ref{sec:conclusions}.

\section{Equation of state, opacities, and neutrinosphere determination}
\label{sec:equat-state-opac}

\subsection{Neutron star model and equation of state}
\label{sec:neutron-star-model}

\begin{figure}[t]
  \centering
  \includegraphics[width=0.9\linewidth]{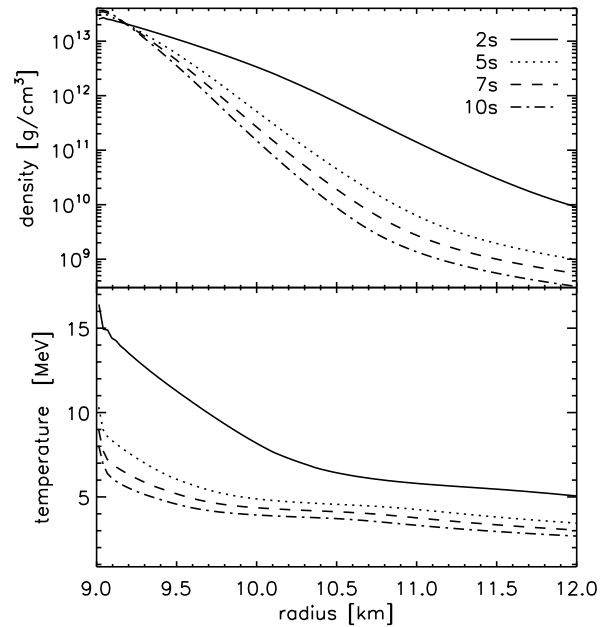}
  \caption{Neutron star atmosphere profiles of density and temperature
    corresponding to model M15-l1-r1 of Ref.~\cite{Arcones.Janka.Scheck:2007}
    for times $t=2, 5, 7$ and $10 \, \text{s}$ post bounce.\label{fig:profile}}
\end{figure}

Our work employs the protoneutron star model M15-l1-r1 of
Ref.~\cite{Arcones.Janka.Scheck:2007}. The model describes the
structure of the surface layers and of the neutrino-driven wind of a
protoneutron star with a baryonic mass of 1.4~M$_\odot$, obtained in a
spherically symmetric simulation of the (parametrized) neutrino-driven
explosion of a 15~M$_\odot$ star.  The thermodynamical state of hot,
dense matter including its baryonic composition is fully characterized
by three independent variables, for example, density, temperature, and
electron chemical potential (or baryon density, energy density, and
electron fraction).  In the supernova context, these have to be
determined by solving the equations of hydrodynamics for the stellar
plasma and the transport equations for neutrinos and antineutrinos of
all flavors, making use of an EOS that relates the pressure and all
other thermodynamic variables to the three basic ones.
Figure~\ref{fig:profile} shows the time evolution of the density and
temperature profiles after core bounce in the region around the
neutrinospheres (``neutron star atmosphere'') as predicted by model
M15-l1-r1. We point out that this model did not include light
elements, therefore the temperature and density are not exactly those
one would obtain if light elements were included.

Using these temperature and density profiles, we can determine the
electron chemical potential (or, equivalently, the electron fraction
$Y_e$) by making the assumption that the matter is in (neutrinoless)
beta equilibrium.  This is fairly well fulfilled in the neutron star
atmosphere at sufficiently late post-bounce times when the
deleptonization of the protoneutron star interior has slowed down and
the nascent neutron star radiates electron neutrinos and antineutrinos
with very similar number luminosities. In this case the chemical
potentials of neutrons, protons, and electrons fulfill the equality
$\mu_n = \mu_p + \mu_e$, which allows one to compute $\mu_e$ by
invoking also charge neutrality, $Y_e = Y_p$, and the relation for the
total baryon number, $Y_p = 1 - Y_n$. Here $Y_p$ and $Y_n$ denote the
number fractions of free plus bound protons and neutrons,
respectively. It is clear that the $Y_e$-profile one thus obtains will
depend on the considered EOS.

\begin{figure}
  \centering
  \includegraphics[width=0.9\linewidth]{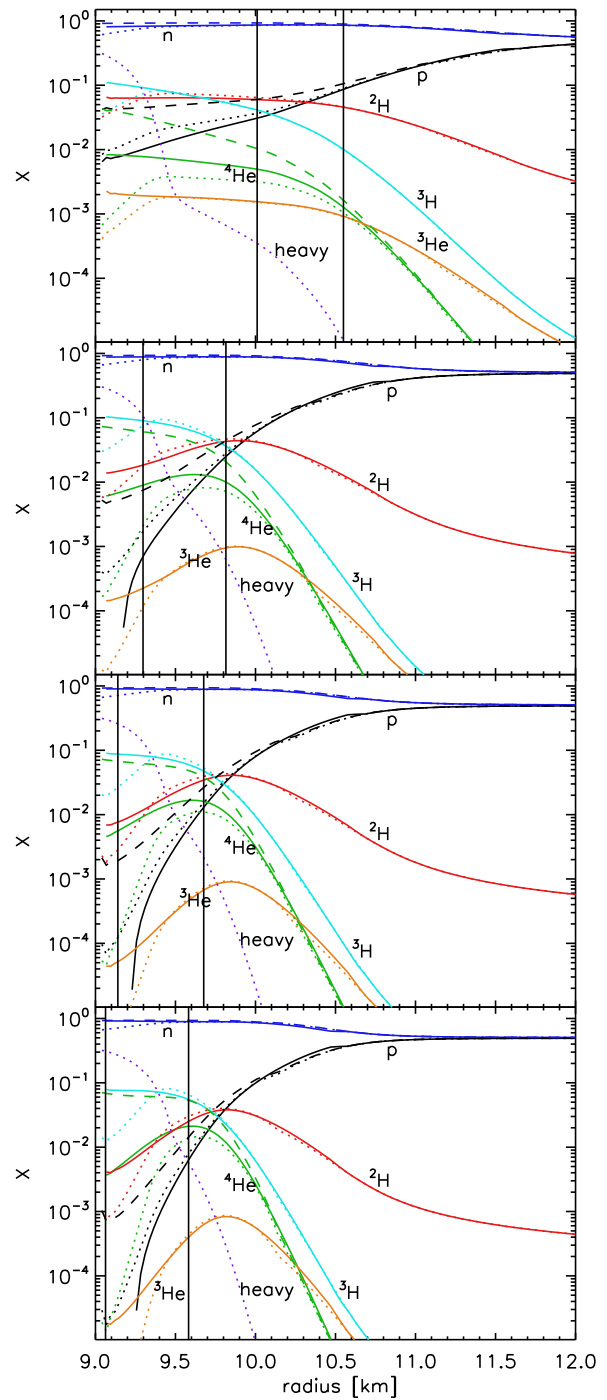}
   \vspace*{-2mm}
  \caption{(Color online) 
    Mass fractions (defined as the mass density of species $i$
    divided by the total mass density) for nuclei present around the
    surface of the protoneutron star using three different EOS: the
    one used in Ref.~\cite{Arcones.Janka.Scheck:2007} (dashed lines),
    the NSE EOS (dotted lines), and the virial
    EOS~\cite{horowitz.schwenk:2006,horowitz.schwenk:2006n,%
oconnor.ea:2008,oconnor.gazit.ea:2007} (solid lines).  
    Beta equilibrium was assumed in each case. The
    lines labeled ``heavy'' represent the mass fraction of nuclei with
    $A > 4$, which are included in the NSE EOS but not in the
    others. From top to bottom the profiles correspond to $t=2, 5, 7$
    and $10 \, \text{s}$ post bounce. The vertical lines mark the
    positions of the neutrinospheres of electron
    neutrinos (right line) and electron antineutrinos 
    (left line).\label{fig:comp}}
\end{figure}

We will compare results obtained with three different EOS. The first
is the one used in Ref.~\cite{Arcones.Janka.Scheck:2007}, which
essentially describes the baryonic composition as a non-interacting
Boltzmann gas of neutrons, protons, and alpha particles in NSE (a
representative heavy nucleus is also included but of no relevance
here).  Calculations performed with this EOS serve as the reference to
which we compare our results.  The second EOS, denoted by NSE EOS,
also assumes matter in NSE but consists of several thousand nuclei,
for which partition functions have been computed in
Ref.~\cite{rauscher:2003}, and includes Coulomb
corrections~\cite{hix.thielemann:1996,Bravo.Garcia-Senz:1999} and
Fermi-Dirac statistics for neutrons and protons.  Our third EOS is the
virial EOS~\cite{horowitz.schwenk:2006,horowitz.schwenk:2006n,%
  oconnor.ea:2008,oconnor.gazit.ea:2007}, which is based on nuclei
with $A \leqslant 4$ and their interactions through second virial
coefficients derived directly from scattering phase shifts.
Figure~\ref{fig:comp} shows the composition obtained from these
different EOS for the thermodynamical conditions of
Fig.~\ref{fig:profile}, assuming beta equilibrium for each of the
displayed cases. In agreement with our arguments, we observe that in
the vicinity of the neutrinospheres the virial and NSE EOS predict
deuteron and triton mass fractions significantly larger than the one
of free protons.

The virial and NSE EOS lead to very similar compositions up to
densities $\rho \sim 10^{13} \gcq$. At these densities, the
treatment of nuclear interactions in the virial EOS becomes
unreliable, and in the NSE EOS interactions are neglected.
This is signaled by a sudden increase in the abundance
of heavy nuclei, when using the NSE EOS, or by a sudden
drop of the proton mass fraction in the virial EOS, due to the
breakdown of the virial expansion with only second virial
coefficients. For densities lower than $\rho \sim 10^{13} \gcq$,
we observe that the main differences of the virial and NSE
EOS are in the alpha particle mass fractions due to attractive
nucleon-alpha interactions~\cite{horowitz.schwenk:2006}.

\subsection{Neutrino opacities and neutrinospheres}
\label{sec:neutr-opac-neutr}

To determine the neutrinosphere radius we
follow Refs.~\cite{keil.raffelt.janka:2003,Shapiro.Teukolsky:1983} and
define the effective neutrino opacity for energy exchange or
thermalization by
\begin{equation}
  \kappa_{\text{eff}} = \sqrt{\kappa_{\text{abs}}
    (\kappa_{\text{abs}} + \kappa_{\text{scatt}})} \,.
  \label{eq:keff}
\end{equation}
The absorption opacity $\kappa_{\text{abs}}$ is considered to include
all processes in which neutrinos exchange energy with the stellar
medium, while the scattering opacity $\kappa_{\text{scatt}}$ contains
those processes where mostly the momentum of the neutrinos is changed
but essentially not their energy. These opacities, $\kappa=\sum n_i
\sigma_i$, are obtained from the neutrino cross sections $\sigma_i$
and number densities $n_i$ of the target particles in the stellar
plasma.

In order to determine an average neutrinosphere radius, rather than an
energy-dependent one, we assume neutrinos to be in thermal
equilibrium with matter up to their so-called average energy sphere,
where the bulk of the neutrino spectrum begins to decouple thermally from
the background medium of the star. We will consider this energy sphere
as the appropriate neutrinosphere in the context of the work presented
in this paper. Until this location the neutrino phase-space distribution
will be assumed to be a Fermi-Dirac distribution function for the local gas 
temperature. Outside of the neutrinosphere the spectral temperature 
is taken to be fixed to its value at the neutrinosphere. Thus,
we average all opacities over the relevant neutrino spectral
distributions and define:
\begin{equation}
  \label{eq:kave}
   \langle \kappa_{\text{eff}} \rangle = \sqrt{\langle
    \kappa_{\text{abs}}\rangle \, (\langle \kappa_{\text{abs}}\rangle  +
    \langle \kappa_{\text{scatt}}\rangle)} \,.
\end{equation}

In the calculation of $\kappa_{\text{scatt}}$ we include elastic
scattering off nucleons~\cite{Horowitz:2002} and
nuclei~\cite{Burrows.Reddy.Thompson:2006} for both electron neutrinos
and antineutrinos. For determining $\kappa_{\text{abs}}$ for electron
neutrinos, it is sufficient to consider their absorption on neutrons,
because neutrons dominate the composition by far. In the case of
electron antineutrinos, besides weak processes~\cite{Horowitz:2002}
with the rare protons we have to include also inverse nucleon-nucleon
bremsstrahlung~\cite{Hannestad.Raffelt:1998} and in particular
charged-current and neutral-current interactions with
deuterons~\cite{Nakamura.Sato.ea:2002} and tritons (as discussed
below). The situation is different for $^3$He as target particle. Its
abundance is always very low so that this contribution to the
antineutrino opacity can be neglected. Moreover, inelastic scattering
on electrons and positrons and neutrino-antineutrino annihilation can
be neglected, because either the cross sections or target densities of
these reactions are typically smaller than those of the neutrino
interactions with baryonic targets.  Ignoring these processes has
hardly any influence on the relative changes of the neutrinospheric
positions that we intend to discuss in this paper. Our approach for
estimating the influence of composition effects in the surface layers
of nascent neutron stars on the position of neutrino-matter decoupling
is rather qualitative and approximative anyway. It is certainly not
suitable for making exact quantitative predictions, a goal that
definitely requires radiation-hydrodynamics simulations with
energy-dependent neutrino transport.

\begin{table}[t]
  \centering
  \caption{Cross sections for charged-current and neutral-current
    antineutrino scattering off $^3$H as a function of incident
    antineutrino energy $E$.\label{tab:nucross}}
  \begin{ruledtabular}
    \begin{tabular}{ccc}
      E [MeV] & $^3$H$(\bar{\nu}_e,e^+)$ [$10^{-42} \, \text{cm}^2$] & 
      $^3$H$(\bar{\nu}_e,\bar{\nu}_e^\prime)$ 
      [$10^{-42} \, \text{cm}^2$] \\ \hline
      11 & $1.66 \times 10^{-5}$ & $2.64 \times 10^{-5}$ \\ 
      12 & $9.56 \times 10^{-5}$ & $4.30 \times 10^{-4}$ \\ 
      13 & $2.40 \times 10^{-4}$ & $1.69 \times 10^{-3}$ \\ 
      14 & $4.49 \times 10^{-4}$ & $4.32 \times 10^{-3}$ \\ 
      15 & $1.79 \times 10^{-3}$ & $8.90 \times 10^{-3}$ \\ 
      16 & $4.97 \times 10^{-3}$ & $1.60 \times 10^{-2}$ \\ 
      17 & $1.13 \times 10^{-2}$ & $2.64 \times 10^{-2}$ \\ 
      18 & $2.19 \times 10^{-2}$ & $4.08 \times 10^{-2}$ \\ 
      19 & $3.80 \times 10^{-2}$ & $6.01 \times 10^{-2}$ \\ 
      20 & $6.13 \times 10^{-2}$ & $8.52 \times 10^{-2}$ \\ 
      21 & $9.32 \times 10^{-2}$ & $1.17 \times 10^{-1}$ \\ 
      22 & $1.35 \times 10^{-1}$ & $1.57 \times 10^{-1}$ \\ 
      23 & $1.90 \times 10^{-1}$ & $2.05 \times 10^{-1}$ \\ 
      24 & $2.58 \times 10^{-1}$ & $2.64 \times 10^{-1}$ \\ 
      25 & $3.43 \times 10^{-1}$ & $3.34 \times 10^{-1}$ \\ 
      26 & $4.46 \times 10^{-1}$ & $4.17 \times 10^{-1}$ \\ 
      27 & $5.70 \times 10^{-1}$ & $5.14 \times 10^{-1}$ \\ 
      28 & $7.18 \times 10^{-1}$ & $6.26 \times 10^{-1}$ \\ 
      29 & $8.93 \times 10^{-1}$ & $7.54 \times 10^{-1}$ \\ 
      30 & $1.10 \times 10^{0}$ & $ 9.01 \times 10^{-1}$ \\ 
      32 & $1.61 \times 10^{0}$ & $ 1.25 \times 10^{0}$ \\ 
      34 & $2.28 \times 10^{0}$ & $ 1.70 \times 10^{0}$ \\ 
      36 & $3.13 \times 10^{0}$ & $ 2.24 \times 10^{0}$ \\ 
      38 & $4.20 \times 10^{0}$ & $ 2.89 \times 10^{0}$ \\ 
      40 & $5.50 \times 10^{0}$ & $ 3.66 \times 10^{0}$ \\ 
      42 & $7.08 \times 10^{0}$ & $ 4.56 \times 10^{0}$ \\ 
      44 & $8.97 \times 10^{0}$ & $ 5.60 \times 10^{0}$ \\ 
      46 & $1.12 \times 10^{1}$ & $ 6.78 \times 10^{0}$ \\ 
      48 & $1.38 \times 10^{1}$ & $ 8.13 \times 10^{0}$ \\ 
      50 & $1.69 \times 10^{1}$ & $ 9.63 \times 10^{0}$ \\ 
      54 & $2.43 \times 10^{1}$ & $ 1.31 \times 10^{1}$ \\ 
      58 & $3.38 \times 10^{1}$ & $ 1.74 \times 10^{1}$ \\ 
      62 & $4.56 \times 10^{1}$ & $ 2.23 \times 10^{1}$ \\ 
      66 & $5.99 \times 10^{1}$ & $ 2.81 \times 10^{1}$ \\ 
      70 & $7.69 \times 10^{1}$ & $ 3.45 \times 10^{1}$ \\ 
      74 & $9.65 \times 10^{1}$ & $ 4.17 \times 10^{1}$ \\ 
      78 & $1.18 \times 10^{2}$ & $ 4.96 \times 10^{1}$ \\ 
      82 & $1.42 \times 10^{2}$ & $ 5.82 \times 10^{1}$ \\ 
      86 & $1.69 \times 10^{2}$ & $ 6.75 \times 10^{1}$ \\ 
      90 & $1.98 \times 10^{2}$ & $ 7.74 \times 10^{1}$ \\ 
      95 & $2.37 \times 10^{2}$ & $ 9.05 \times 10^{1}$ \\ 
     100 & $2.80 \times 10^{2}$ & $ 1.04 \times 10^{2}$ 
\end{tabular}
\end{ruledtabular}
\end{table}

The first {\it ab-initio} calculations for neutral-current inelastic
cross section off $^3$H and $^3$He nuclei were presented in
Ref.~\cite{oconnor.gazit.ea:2007}, where the neutrino energy
was averaged over a Fermi-Dirac spectrum for given temperature.
For our purpose, it is advantageous to have the total cross sections 
as a function of neutrino energy. In addition, charged-current
cross sections for antineutrinos on tritons are needed. Therefore,
we have computed the relevant cross sections using a model based
on the random phase approximation (RPA)
which has been successfully applied to the study of many
neutrino-induced reactions (for example, see 
Ref.~\cite{Kolbe.Langanke:2001,Kolbe.Langanke.ea:2003}). Our approach
follows the one described in Ref.~\cite{Kolbe.Langanke.Vogel:1999},
where we adopt an RPA that distinguishes between proton and neutron
degrees of freedom for the particle and hole states. The parent ground
state is approximated by the lowest independent particle model state
with the single-particle energies derived from an appropriate
Woods-Saxon potential that reproduces the particle separation
energies in the parent nucleus. The partial occupancy formalism as
described in Ref.~\cite{Kolbe.Langanke.Vogel:1999} is applied to the
proton and neutron states and holes for $^3$H and $^3$He, respectively.
As residual interaction, we have used the Landau-Migdal force of
Ref.~\cite{Kolbe.Langanke.Vogel:1999}. The calculation includes all
multipole transitions with $\lambda \leqslant 4$ and both parities, properly
accounting for the dependence of the multipole operators on the
momentum transfer~\cite{Walecka:1975,Donnelly.Peccei:1979}. The
Gamow-Teller strength has been quenched by a factor $0.74$ as in
shell-model calculations~\cite{Martinez-Pinedo.Poves.ea:1996b}. Our
calculated cross sections are listed in Table~\ref{tab:nucross} as a
function of incident antineutrino energy $E$. To compare our results with
the {\it ab-initio} calculations of Ref.~\cite{oconnor.gazit.ea:2007},
we have folded the cross sections with a Fermi-Dirac neutrino 
distribution and find agreement to better than a few percent in all cases.

From the effective opacity of Eq.~(\ref{eq:kave}), we determine the
optical depth as:
\begin{equation}
  \label{eq:odepth}
  \tau(r) = \int_r^\infty \langle \kappa_{\text{eff}} \rangle \, dr \,.
\end{equation}
The neutrinospheric radius $R_\nu$ is then defined as the position
where the optical depth reaches $\tau(R_\nu) =
2/3$~\cite{Mihalas:1978}.
We assume that the neutrino distribution function at the neutrinosphere
is represented by a Fermi-Dirac spectrum of temperature $T$ and effective
degeneracy parameter $\eta_\nu$:
\begin{equation}
  \frac{dn_\nu}{dE}(E) = \frac{1}{2\pi^2(\hbar c)^3} 
    \frac{E^2}{\exp[E/(k T_\nu) -
    \eta_\nu] + 1} \,. 
\end{equation}
The neutrino number luminosity~\cite{Janka:1991} at the neutrinosphere
is then given by
\begin{equation}
  \label{eq:lumi}
  L_{n,\nu} = \frac{2 c}{\pi (\hbar c)^3} \, (k T_\nu)^3 \, R_\nu^2 \,
  f(R_\nu) \, F_2(\eta_\nu) \,,
\end{equation}
and a similar equation for antineutrinos. Here $f$ is the flux
factor needed to convert number density to number flux, with 
$f(R_\nu) \approx 0.25$ at the location of the 
neutrinosphere~\cite{Janka:1991}.
The Fermi integral for relativistic particles is defined as
$F_n (\eta) = \int_0^\infty
dx \, x^n/[1+\exp(x-\eta)]$, where $\eta$ should be regarded as a
spectral parameter that is not necessarily related to the neutrino
degeneracy.

The above procedure allows for a completely independent determination
of the electron neutrino and antineutrino spheres, and
consequently of their number luminosities. However, the number
luminosities are constrained by the requirement that the net flux of
electron neutrinos minus electron antineutrinos
is positive or zero to warrant that the lepton content of the
star does not increase. Simulations~\cite{Pons.Reddy.ea:1999}
show that the nascent neutron star deleptonizes only very gradually
and therefore the number luminosities of neutrinos and antineutrinos are
rather similar, allowing us to make the approximation 
that the net flux is equal to zero,
$L_{n,\nu} = L_{n,\bar{\nu}}$. This choice is consistent 
with our assumption of beta equilibrium for fixing the composition
in the outer layers of the neutron star
(see Sect.~\ref{sec:neutron-star-model}), 
in which case the same numbers
of electron neutrinos and antineutrinos are created in this region. 
Since our calculations are 
mainly sensitive to the difference in the spectral properties
of neutrinos and antineutrinos, we set $\eta_{\bar{\nu}} = 0$ 
for simplicity and
obtain $\eta_\nu$ from the condition of zero net flux.

We use the following iterative procedure~\cite{Keil.Janka:1995} 
to determine the $\nu_e$ and
$\bar{\nu}_e$ neutrinospheres. First, we assume some initial values
for the temperatures of the neutrino and antineutrino distributions,
and determine $\eta_\nu$ from the condition of zero net flux. Using
these values we compute the radii, at which the optical depths become
2/3. Then, we set the new $\nu_e$ and $\bar{\nu}_e$ temperatures to
the local temperature at the corresponding radii and update
$\eta_\nu$ from the zero net flux condition. We iterate this procedure
until convergence.

\begin{figure}
  \centering
  \includegraphics[width=0.9\linewidth]{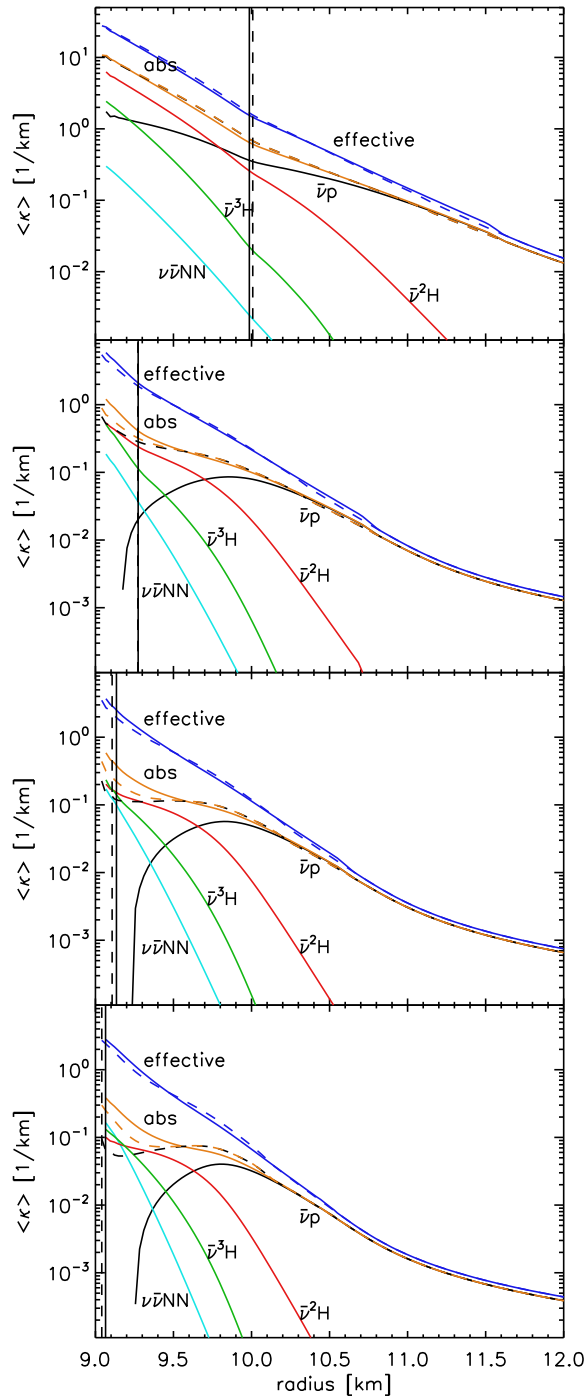}
  \caption{(Color online)
    Different contributions to the absorption opacity
    $\langle\kappa_{\text{abs}}\rangle$ 
    and effective opacity $\langle\kappa_{\text{eff}}\rangle$ of
    electron antineutrinos 
    as a function of radius for $t = 2, 5, 7$ and $10 \,
    \text{s}$ post bounce (from top to bottom) and for two different EOS:
    the one used in Ref.~\cite{Arcones.Janka.Scheck:2007} (dashed lines)
    and the virial EOS (solid lines), based on the target abundances displayed
    in Fig.~\ref{fig:comp}. In addition, the total absorption
    (``abs'') and total effective opacity (``effective'') are shown. 
    The vertical lines mark the positions of the neutrinospheres of electron 
    antineutrinos for the two different EOS.
    In the second panel from the top the vertical dashed and 
    solid lines coincide.\label{fig:tau}}
\end{figure}

Since mainly antineutrinos are affected by the changes in composition,
we focus in Fig.~\ref{fig:tau} on the different contributions to the
antineutrino absorption opacity $\langle \kappa_{\text{abs}}\rangle$
and the effective opacity $\langle\kappa_{\text{eff}}\rangle$, at four
different times post bounce. The dashed lines represent our reference
calculation, which considers only antineutrino absorption on protons,
nucleon-nucleon bremsstrahlung and neglects the presence of light
nuclei, corresponding to the treatment in
Ref.~\cite{Arcones.Janka.Scheck:2007}.  For comparison, the solid
lines are based on the virial EOS composition and the antineutrino
interactions as discussed above (similar results are obtained using
the NSE EOS). We find in Fig.~\ref{fig:tau} that the antineutrino
$\langle\kappa_{\text{abs}}\rangle$ in the region around the
neutrinosphere is dominated by the contribution from deuterons and
tritons, for which charged-current and neutral-current reactions
contribute approximately in equal amounts.  Although tritons become
more abundant (see Fig.~\ref{fig:comp}), deuterons become more
important for the opacities.

\section{Neutrino luminosities, spectra, and neutron excess
in the ejecta}
\label{sec:neutr-lumin-spectra}

In the following, we explore the influence of light nuclei on the
properties of the emitted neutrinos. To this end, we consider four
different cases denoted A to D in Table~\ref{tab:models}, which use
the different EOS discussed above.  Our reference calculation is
case~A, which is performed with the same EOS (consisting of neutrons,
protons, and $\alpha$-particles in NSE) and the same neutrino
reactions as in Ref.~\cite{Arcones.Janka.Scheck:2007}. When we change
to the improved EOS including light nuclei for given density,
temperature, and $Y_e$, the different baryonic composition leads to a
modified optical depth and therefore to a shift of mainly the position
of the neutrinosphere of electron antineutrinos. On the other hand,
for given density and temperature in the neutron star atmosphere, the
improved EOS imply different nucleon chemical potentials and therefore
yield a different value of $Y_e$ when we impose the constraint of beta
equilibrium for the stellar matter (see
Sect.~\ref{sec:neutron-star-model}).  This again influences the
composition, optical depth, and neutrinospheric positions.  To
quantify separately the impact due to the direct compostion change and
the one associated with an adjustment to a new beta equilibrium, we
define an intermediate case~B that uses the same density, temperature,
and $Y_e$ profiles as case~A but obtains the baryonic composition from
the improved NSE EOS.  Case~C then uses the same EOS as case~B but
takes into account the adjustment to a new beta equilibrium.  In
case~D, the baryonic composition as well as the beta equilibrium state
are based on the virial EOS.

Table~\ref{tab:rtye} shows the neutrinosphere radii and
corresponding properties of electron neutrinos and antineutrinos
for the four different 
cases. The average energies, defined as
\begin{equation}
\langle \epsilon_{\nu} \rangle = \frac{F_3(\eta_\nu)}{F_2(\eta_\nu)} \, k T_\nu \,,
\end{equation}
and luminosities computed for our
reference case~A are in good agreement with the results of
Ref.~\cite{Arcones.Janka.Scheck:2007}. In order to compare the
average energies and luminosities of Table~\ref{tab:rtye} with
those commonly used in nucleosynthesis studies (which correspond
to values measured at infinity), the gravitational redshift must
be included. For our neutron star model, the redshift correction
corresponds to a reduction by a factor of about 0.8.

\begin{table}[t]
  \caption{Different cases explored in Sect.~\ref{sec:neutr-lumin-spectra}.
    \label{tab:models}} 
  \begin{ruledtabular}
    \begin{tabular}{ccc}
      Case & $Y_e$ determined from & EOS and composition \\ \hline
      A & beta equilibrium &  NSE (n, p, $^4$He) \\
      B & case A &  NSE (nucleons and nuclei) \\ 
      C & beta equilibrium &  NSE (nucleons and nuclei) \\  
      D & beta equilibrium &  virial (n, p, $A \leqslant 4$ nuclei) 
    \end{tabular}
  \end{ruledtabular}
\end{table}

\begin{figure}
  \centering
  \includegraphics[width=0.9\linewidth]{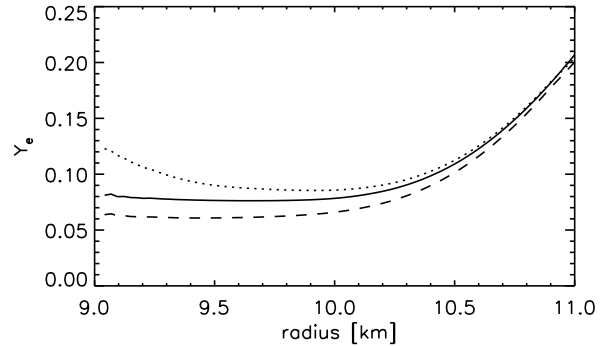}
  \caption{Profile of the electron fraction $Y_e$ in the region around the
    neutrinosphere at $t=2 \, \text{s}$ after core
    bounce. The dashed line corresponds to the EOS of
    Ref.~\cite{Arcones.Janka.Scheck:2007} (case~A), the dotted line to the NSE EOS (case~C),
    and the solid line to the virial EOS (case~D). All profiles are obtained
    assuming beta equilibrium for the corresponding EOS.\label{fig:yeprof}}
\end{figure}

\begin{table*}[t]
  \caption{Neutrinosphere radii $R_{\bar{\nu}_e,\nu_e}$,
    neutrino spectral temperatures $T_{\bar{\nu}_e,\nu_e}$, and
    average energies $\langle \epsilon_{\bar{\nu}_e,\nu_e} \rangle$,
    as well as number luminosities $L_n$, spectral parameter $\eta_{\nu_e}$,
    and wind electron fractions $Y_e^{\text{w}}$ at four different
    times post bounce.\label{tab:rtye}}
  \begin{ruledtabular}
  \begin{tabular}{lccccccccc}
& $R_{\bar{\nu}_e}$ & $T_{\bar{\nu}_e}$ & $\langle \epsilon_{\bar{\nu}_e} \rangle$
& $L_n$ & $\eta_{\nu_e}$ & $R_{\nu_e}$ & $T_{\nu_e}$ & 
$\langle \epsilon_{\nu_e} \rangle $ & $Y_e^{\text{w}}$ \\
& [km] & [MeV] & [MeV] &  [$10^{56} \, \text{s}^{-1}$] &  & 
[km] & [MeV] & [MeV] & \\ \hline
\multicolumn{10}{c}{$t = 2 \, \text{s}$} \\ \hline
A & 10.01 & 8.14 & 25.64 & 6.05 & 0.72 & 10.55 & 6.34 & 20.71 & 0.514 \\ 
B &  9.977 & 8.30 & 26.16 & 6.38 & 0.79 & 10.55 & 6.34 & 20.80 & 0.507 \\ 
C & 10.00 & 8.17 & 25.73 & 6.10 & 0.73 & 10.55 & 6.35 & 20.75 & 0.513 \\ 
D &  9.979 & 8.29 & 26.12 & 6.36 & 0.77 & 10.53 & 6.37 & 20.87 & 0.509 \\ \hline
\multicolumn{10}{c}{$t = 5 \, \text{s}$} \\ \hline
A & 9.272 & 7.17 & 22.60 & 3.55 & 1.01 & 9.821 & 5.14 & 17.10 & 0.478 \\ 
B & 9.260 & 7.24 & 22.83 & 3.65 & 1.04 & 9.819 & 5.15 & 17.16 & 0.475 \\ 
C & 9.295 & 7.04 & 22.17 & 3.37 & 0.94 & 9.814 & 5.16 & 17.07 & 0.487 \\ 
D & 9.272 & 7.17 & 22.60 & 3.55 & 1.00 & 9.813 & 5.16 & 17.15 & 0.480 \\ \hline
\multicolumn{10}{c}{$t = 7 \, \text{s}$} \\ \hline
A & 9.107 & 6.88 & 21.69 & 3.03 & 1.15 & 9.683 & 4.73 & 15.90 & 0.462 \\ 
B & 9.095 & 6.97 & 21.95 & 3.13 & 1.19 & 9.681 & 4.74 & 15.96 & 0.458 \\ 
C & 9.139 & 6.68 & 21.04 & 2.78 & 1.04 & 9.676 & 4.75 & 15.82 & 0.475 \\ 
D & 9.134 & 6.71 & 21.14 & 2.82 & 1.05 & 9.675 & 4.75 & 15.85 & 0.473 \\ \hline
\multicolumn{10}{c}{$t = 10 \, \text{s}$} \\ \hline
A & 9.041 & 6.94 & 21.86 & 3.06 & 1.49 & 9.592 & 4.37 & 15.05 & 0.431 \\ 
B & 9.039 & 7.02 & 22.12 & 3.17 & 1.53 & 9.590 & 4.37 & 15.12 & 0.427 \\ 
C & 9.063 & 6.49 & 20.44 & 2.51 & 1.23 & 9.582 & 4.39 & 14.82 & 0.456 \\ 
D & 9.065 & 6.45 & 20.32 & 2.47 & 1.20 & 9.581 & 4.39 & 14.80 & 0.458 \\ 
\end{tabular}
\end{ruledtabular}
\end{table*}

The differences in the various observables of
Table~\ref{tab:rtye} caused by the presence of light nuclei 
can be understood from a detailed comparison of the four cases.
While case~A considers the neutron star
atmosphere to be in beta equilibrium with only neutrons, protons, and alpha
particles, case~B uses the same $Y_e$ profile as case~A,
but accounts for the presence of light nuclei. Their appearance
happens at the expense of the number of free protons, whose mass
fraction is drastically reduced. This leads to a lower
antineutrino opacity than in case~A, even when the additional 
antineutrino interactions with light nuclei are fully included.
Consequently, the neutrinosphere of antineutrinos moves to a smaller
radius where the temperature is larger so that antineutrinos are 
expected to be radiated with a higher mean energy.   

However, as discussed above, the EOS with light nuclei also lead to a
shift of the beta equilibrium conditions in the neutron star
atmosphere. The balance of electron neutrino and antineutrino
absorption and production reactions is locally established for a
higher value of $Y_e$, thus compensating for the reduced abundance of
free protons in the presence of light nuclei. This effect is shown in
Fig.~\ref{fig:yeprof} for the $Y_e$ profile at $t=2 \, \text{s}$ from
the different EOS. The shift of the beta equilibrium is taken into
account in cases~C and D. As a consequence of the higher $Y_e$, the
abundances of protons, deuterons, and tritons are larger than in
case~B (although free protons are clearly reduced relative to
their abundance in case~A, see Fig.~\ref{fig:comp}).  Compared to
case~B, the antineutrino opacity is therefore increased and the
corresponding neutrinosphere is located at a larger radius, leading to
a lower mean energy of the escaping antineutrinos.

In contrast to the properties of electron antineutrinos, the spectra
of electron neutrinos are only slightly affected by the improved EOS
and by the adjustment to a new beta equilibrium. There are two reasons
for this: The mass fraction of free neutrons dominates the composition
and only slightly differs for all cases, and second, weak reactions
with neutrons are responsible for most of the opacity of electron
neutrinos.

From this general discussion we conclude that the change of the
baryonic composition due to light nuclei on the one hand and due to the
neutrino-driven adjustment of matter to a new beta equilibrium on the
other have effects in opposite directions for the position of the
antineutrinosphere (and rather unimportant effects on the
neutrinosphere).  This makes it difficult to reliably predict the
change of this position in cases~C and D compared to the reference
case~A. In fact, as Table~\ref{tab:rtye} shows, the outcome of these
competing effects can go either way.  At early times ($t=2 \, \text{s}$
after bounce) the matter in the vicinity of the neutrinosphere has
large temperatures and hence a rather large fraction of free protons
is present (see Fig.~\ref{fig:comp}).  Under such conditions protons
dominate the antineutrino opacity in the region around the
antineutrinosphere (see Fig.~\ref{fig:tau}).  In cases~C and D the
proton mass fraction is noticeably reduced, but the additional opacity
contributions due to light nuclei cannot compensate for the reduction
of the proton opacity. As a result the antineutrinosphere moves to
slightly smaller radii resulting in larger average energies for the
emitted antineutrinos. As the protoneutron star and its atmosphere
cool, the surface density profile steepens (see
Fig.~\ref{fig:profile}), and the neutrinospheres move to smaller radii
within the same model. The matter in the region of the neutrinospheres
becomes more neutron-rich and the proton abundance is lower. However,
in cases~C and D substantial amounts of deuterons and tritons are
present in this region overcompensating the reduction in the proton
mass fraction and making the total antineutrino absorption opacities
higher in cases~C and D compared to case~A. Consequently, the
antineutrinosphere moves to larger radii resulting in smaller
antineutrino average energies.

We emphasize that at late times the antineutrinosphere is located at
densities $\rho \gtrsim 10^{13} \gcq$, where nuclear interactions and
many-body contributions affect the composition and neutrino cross
sections. For example, for $t = 7$ and $10 \, \text{s}$, the
contribution from densities $\rho > 10^{13} \gcq$ to the optical depth
is 52\% and 65\%, respectively. In the present work, as well as in
state-of-the-art studies of protoneutron star
winds~\cite{Arcones.Janka.Scheck:2007}, such potentially important
effects have been neglected. They should be considered in future
work.

In order to quantify the effects of the changing antineutrino 
energies on the nucleosynthesis conditions in the baryonic wind
driven by neutrino energy deposition off the neutron star surface,
we estimate the wind electron fraction $Y_e^{\text{w}}$
from the expression~\cite{Qian:2003}
\begin{equation}
  \label{eq:ye}
  Y_e^{\text{w}} = \frac{\lambda_{\nu_e n}}{\lambda_{\nu_e n} +
    \lambda_{\bar{\nu}_e p}} \,.
\end{equation}
Here $\lambda_{\nu_e n}$ and $\lambda_{\bar{\nu}_e p}$ are the
neutrino absorption rate on neutrons and the antineutrino absorption
rate on protons, respectively, which depend on the
neutrino number luminosities
$L_{n,\nu}$, the neutrino spectra, and the radial distance $r$ 
from the neutron star:
$\lambda_{\nu} = L_{n,\nu} \, \langle \sigma \rangle
/[4 \pi r^2 f(r)]$,
where $\langle \sigma \rangle$ is the relevant cross section
suitably averaged over the neutrino 
spectrum~\cite{Qian:2003,Qian.Woosley:1996}
and including weak magnetism corrections~\cite{Horowitz:2002}. 
In making use of Eq.~(\ref{eq:ye}), we assume that the ejected
matter is initially composed only of neutrons and protons, so
we neglect the presence of alpha particles and
the so-called alpha-effect~\cite{Meyer.Mclaughlin.Fuller:1998}.   
Furthermore, we suppose that the matter is exposed 
long enough to neutrino and antineutrino captures to achieve an equilibrium 
between neutrino and antineutrino absorptions, and that this
happens at such large distances (and low temperatures) that 
electron and positron captures can be ignored.
 
Using Eq.~(\ref{eq:ye}), we have calculated the wind electron
fraction $Y_e^{\text{w}}$ for the different cases. The corresponding
results are given in Table~\ref{tab:rtye}.
Because of the increase of the average antineutrino energies,
we find that matter is ejected in cases~C and D at early times
with slightly lower $Y_e^{\text{w}}$ values than in case~A.
At later times ($t \gtrsim 5 \, \text{s}$ after bounce) the
mean antineutrino energies are smaller in cases~C and D than in
case~A, and therefore $Y_e^{\text{w}}$ is slightly higher than in
our reference case.

Once the ejected matter has reached larger radii and thus
low enough temperatures, nuclei can form.
During the early phases of nucleosynthesis mainly alpha particles
but also light nuclei are present in the composition. Neutrino
interactions with these light elements can constitute an additional
source of energy deposition in the wind. As noted by Qian and
Woosley~\cite{Qian.Woosley:1996} such an additional source of energy 
could
increase the entropy and reduce the dynamical time scale of the wind,
and consequently facilitate the production of heavy nuclei via the
r-process~\cite{hoffman.woosley.qian:1997}. Using one of the wind
trajectories resulting from the hydrodynamical simulations of
Ref.~\cite{Arcones.Janka.Scheck:2007}, we have computed the contribution
to the energy deposition rate arising from light nuclei (deuterons,
tritons, and alpha particles). Among the light nuclei, we found that
the dominating contribution
comes from neutrino interactions with alpha particles (where
we have taken the cross sections from Gazit
and Barnea~\cite{Gazit.Barnea:2007}). However, our calculation 
showed that the additional
energy deposition provided by light elements is too low, by more than
an order of magnitude, to have an impact on the conditions
for r-process nucleosynthesis.

\section{Conclusions}
\label{sec:conclusions}

We have shown that the thermodynamical conditions in the outer
layers of a protoneutron star favor the presence of light nuclei,
mainly deuterons and tritons, which are not accounted
for by EOS currently used in core-collapse supernova
simulations~\cite{Janka.Langanke.ea:2007} and in studies of 
neutrino-driven
supernova outflows~\cite{Arcones.Janka.Scheck:2007}. Using the
profiles of a hydrodynamical model for
neutrino-driven supernova ejecta~\cite{Arcones.Janka.Scheck:2007},
we have estimated the effects of light nuclei on
the emission of electron neutrinos and antineutrinos.
For this purpose we have compared the virial 
and NSE EOS, which include light nuclei, to a reference case
composed of neutrons, protons, and alpha particles.

The abundance of light nuclei can be studied in laboratory heavy-ion
collisions. These experiments can reproduce the densities and 
temperatures near the neutrinosphere. In the future, because the 
neutrinosphere is neutron rich, the abundances of light nuclei 
should be measured for more neutron-rich systems. This can be 
done with radioactive beams.

The appearance of light nuclei has only a minor
impact on the position of the electron neutrinosphere and consequently
on the average energy of the radiated electron neutrinos.
However, the situation is different for
electron antineutrinos. At early times when the protoneutron star is
relatively hot and protons have mass fractions around 0.1 in the
neutrinospheric region, the
appearance of light nuclei reduces the antineutrino
opacity. Therefore, antineutrinos escape from hotter layers
in the protoneutron star with slightly larger average energy. At
later times the mass fraction of protons in the protoneutron star is
greatly reduced, but light elements (in particular tritons) can have
mass fractions that reach values even around 0.1.
This makes light nuclei the
major source of opacity for antineutrinos. Comparing with the EOS
used in Ref.~\cite{Arcones.Janka.Scheck:2007}, we see that antineutrinos
are kept in thermal equilibrium with matter until larger radii in the
protoneutron star, reducing the average energy of the emitted
antineutrinos. For the latest time considered in the present study
($t=10 \, \text{s}$ after bounce), the reduction could be as large
as $1.5 \, \mev$.

The changes in the antineutrino average energies can have consequences
for the nucleosynthesis occurring in neutrino-driven winds. Such
winds are a very interesting
nucleosynthesis site. They allow for proton-rich
ejecta during the first couple of
seconds~\cite{Pruet.Woosley.ea:2005,Froehlich.Hauser.ea:2006},
where the recently suggested $\nu
p$-process~\cite{Froehlich.Martinez-Pinedo.ea:2006,Pruet.Hoffman.ea:2006}
may take place, as well as
neutron-rich ejecta at later times, which might provide the 
conditions for r-process
nucleosynthesis~\cite{arnould.goriely.takahashi:2007}. During the early
proton-rich phase, the changes in the antineutrino energies are minor
and consequently the electron-to-baryon ratio $Y_e^{\text{w}}$ in the wind
remains the same or is slightly reduced. For the neutron-rich phase 
and in particular for the latest
times considered in the present study, 
we find that $Y_e^{\text{w}}$ can increase
by as much as 0.025. This is a substantial change, and if everything
else remains the same, such a change makes the occurrence of 
strong r-processing less likely.

For more detailed studies, reliable estimates of the neutrino cross
sections with light nuclei are desirable. The neutrino-deuteron cross
sections~\cite{Nakamura.Sato.ea:2002} used here are probably
sufficiently accurate at the neutrino energies considered in this
work. For charged-current and neutral-current cross sections of
antineutrino reactions with tritons, we have presented results
based on a relatively simple RPA approach. The neutral-current 
cross sections were found to agree very well with the recent
{\it ab-initio} results of Ref.~\cite{oconnor.gazit.ea:2007}. Similar
calculations for charged-current cross sections would be useful.

Our estimates show that future simulations of neutrino-driven
supernova outflows should take into account light elements in the
baryonic composition of the stellar medium and the corresponding cross
sections of neutrino interactions, especially those of electron
antineutrinos.  This is particularly important to fully quantify the
consequences for the properties of the neutrino emission from forming
neutron stars and for the neutrino-generated nucleosynthesis
conditions in the baryonic mass that is lost from such stars.  In
addition to including light nuclei in the EOS, our results show that,
for late times, an important future problem is understanding the
properties of and neutrino interactions with nucleonic matter at
densities above $10^{13} \gcq$.

\begin{acknowledgments}
  A.~Arcones acknowledges support by the Deutsche
  Forschungsgemeinschaft through contract SFB 634 ``Nuclear structure,
  nuclear astrophysics and fundamental experiments at small momentum
  transfers at the S-DALINAC''.  The work of E.~O'Connor and
  A.~Schwenk is supported by the Natural Sciences and Engineering
  Research Council of Canada (NSERC). TRIUMF receives federal funding
  via a contribution agreement through the National Research Council
  of Canada.  The work of H.-T.~Janka is supported by the Deutsche
  Forschungsgemeinschaft through the Transregional Collaborative
  Research Centers SFB/TR~27 ``Neutrinos and Beyond'' and SFB/TR~7
  ``Gravitational Wave Astronomy'', and the Cluster of Excellence
  EXC~153 ``Origin and Structure of the Universe''
  (http://www.universe-cluster.de). The work of C.J.~Horowitz is
  supported in part by DOE grant DE-FG02-87ER40365.
\end{acknowledgments}


\end{document}